\documentclass[aps,prd,groupedaddress,onecolumn,preprint]{revtex4-1}

\usepackage{amssymb,amsmath}
\usepackage{graphicx}
\usepackage{hyperref}

\begin{document}


\title{Generation of Cosmological Flows in General Relativity \\
(Features and Properties of Integrable Singularities)}


\author{V.N. Lukash}
\email{lukash@asc.rssi.ru}

\author{E.V. Mikheeva}
\email{helen@asc.rssi.ru}

\affiliation{Astro Space Centre of the Lebedev Physical Institute
\\ ul. Profsoyuznaya, 84/32, 117997 Moscow, Russia}

\author{V.N. Strokov}
\email{strokov@asc.rssi.ru}

\affiliation{Departamento de F{\'i}sica -- ICE -- Universidade
Federal de Juiz de Fora \\Juiz de Fora, MG, Brasil -- CEP:
36036-330.}

\altaffiliation[On leave from ]{Astro Space Centre of the Lebedev
Physical Institute.}


\date{\today}

\begin{abstract}
We discuss status of the singularity problem in General Relativity
and argue that the requirement that a physical solution must be
completely free of singularities may be too strong. As an example,
we consider properties of the integrable singularities and show
that they represent light horizons separating T-regions of black
and white holes. Connecting an astrophysical black hole to a white
hole, they lead to a natural mechanism of generating new
universes. Under favorable conditions the new universes will also
contain black holes which, in their turn, will give rise to
another generation of universes. In this case the cosmological
evolutionary tree will continue to grow to form the
``hyperverse''. This scenario essentially differs from other known
mechanisms, such as bounce, birth from ``nothing'', baby-universe
scenario, etc.
\end{abstract}

\pacs{04.70.-s, 04.20.Dw, 98.80.Bp}

\maketitle


\section{Introduction}

Singularities in solutions of equations that are supposed to
describe a physical phenomenon are usually associated with some
kind of idealization. For example, in the case of hydrodynamics of
ideal fluids one has multiple solutions in the problem of a body
immersed in a flow~\cite{landafshits-6}, a paradox resolved by
taking into account a non-zero viscosity. As for gravitational
singularities, they already appear in Newtonian gravity. A sphere
of particles with initially equal radial velocities will collapse
to a point with infinite density. However, this singularity
disappears as soon as we introduce random perturbations to the
initial velocities. In particular, effective cutoff of small-scale
perturbations in N-body simulations of dark matter halos may be
the reason~\cite{cusp-problem} why the singular
Navarro--Frenk--White ``cusp'' profile~\cite{NFW} does not
reproduce observations.

In General Relativy (GR) the situation is somewhat different and
the introduction of small perturbations in initial conditions does
not prevent the formation of the black-hole singularity in
quasi-spherical collapse~\cite{price-a,*price-b,*thorne}.
Moreover, as it was recognized by a set of
theorems~\cite{penrose65,*hawkingpenrose}, the singularities in GR
form under a wide range of initial conditions. A belief arose that
it has to do with the idealization that is neglecting
quantum-gravitational effects and, hence, corrections to the
Einstein equations at energies of the order $M_{Pl}\sim
10^{19}$~GeV. Indeed, a consistent treatment of quantum field
theory in curved space--time makes it necessary to introduce
corrections to the Einstein--Hilbert gravitational action
(see~\cite{shapiro} for a review). Possible consequences of such
corrections for the problem of the black-hole singularity was
first studied in~\cite{frolovvilkovsky}. Nevertheless, what is the
exact form of the corrections is unclear until now, and the issue
remains.

The latter fact led to attempts of treating the problem
phenomenologically. That is, all possible corrections are ascribed
to an effective energy--momentum tensor (EMT) and the Einsteinian
form of the equations if preserved. If we assume that the
singularity (later on, we consider the Schwarzschild singularity)
must somehow disappear than, at least near it, the effective EMT
acquires an exotic form. One can go even further and put the
principle of finiteness of curvature and/or density as a universal
nature law~\cite{markov-1,*markov-2}. Particularly, this principle
led to a model~\cite{markovfrolov}, in which the Schwarzschild
geometry is attached to the de Sitter one in the vicinity of
singularity where the squared Rimann tensor
$R_{\alpha\beta\gamma\delta}R^{\alpha\beta\gamma\delta}$ achieves
the Planckian value $\sim M_{Pl}^{4}$. As the inflationary
paradigm states~\cite{guth,*linde}, the early universe had a
quasi-de-Sitter symmetry. Hence, in the cosmological context the
model gives a mechanism of generating a universe in the interior
of a black hole.

The idea that black holes can serve as generators of new universes
was realized a long time ago~\cite{Nov66}. It takes advantage of
two remarkable features. First, a time-reversed solution of the
Einstein equations is also a solution. Thus, in the interior of a
white hole -- the time-reversed counterpart of black hole --
particles can only move away from the singularity, which is
reminiscent of the geometry of an expanding universe. Second, the
collapse into black hole ends with a singularity while our
universe should have begun with a singularity (see~\cite{lm} and
refs. therein), and the two singularities may be, in fact, one.

The phenomenological approach, however, does not permit to define
the equation-of-state of the effective matter. It makes tempting
to consider cases with a high degree of symmetry like that of the
above-mentioned de Sitter space--time. In the meantime, there are
indications that less symmetric structures of the singular region
are possible. First, the black-hole interior exhibits the symmetry
$\mathbb{R}\times\mathbb{S}^{2}$ of the Kantowski--Sachs model.
Second, in approaching to the singularity the
Belinskii--Lifshitz--Khalatnikov regime with alternating Kasner
eras is realized. One may ask the question: what is the advantage
of this if the singularity persists? It does, but its structure
may be less harmful in the sense that in some directions tidal
forces remain finite. If the effective matter that forms the
geometry moves only in these special directions, it can pass the
singularity to form the geometry beyond it. In another context,
this kind of situation happens, for example, in the cosmological
model with mixture of the Chaplygin gas and
dust~\cite{Chap-singularity}.

As applied to the black-hole singularity the described situation
takes places for the class of the so-called integrable
singularities proposed in~\cite{ls,lsm}. The main assumption is
that gravitational potentials in a Schwarzschild-like metric
remain finite. Note that it does not imply the finiteness of the
curvature tensor, though tidal forces on some geodesics are
finite. This allows us to continue the effective-matter flow to
the expanding white-hole region and, thus, realize a natural
mechanism of generating new universes in the interior of a black
hole without the extra assumption on the absence of the
singularity. Since other known scenarios of cosmogenesis (bounce,
birth from ``nothing'', baby universes) use this extra assumption,
our scenario is instrinsically different.

In this paper we further study properties of space--time near the
integrable singularity and generalize previously developed model
of eternal black-white hole for the more realistic case of a
universe born by an astrophysical black hole. The paper is
organized as follow. In Sections~\ref{continuation}
and~\ref{space-time} we discuss the issue of continuation of
Schwarzschild-like metrics through the integrable singularity and
study spacetime structure near it. In Section~\ref{geodesics} we
classify geodesic behavior in its vicinity.
Section~\ref{black-white} is a r{e'}sum{e'} of the model of
black-white hole while in Section~\ref{astrophysical} a model of
an {\it astrogenic} (formed inside a collapsed astrophysical black
hole) universe is presented. In Section~\ref{nature} we discuss
possible physical nature of integrable singularities and in
Section~\ref{arrow} -- their role in the global picture of
cosmogenesis. In Conclusion we briefly summarize basic properties
of astrogenic universes.

\section{How to continue the Schwarzschild metric}\label{continuation}

A black hole with positive external mass $M>0$ without rotation
and charge is described in GR by the Schwarzschild metric in
vacuum:
\begin{equation}
\label{schwarz}
ds^2=\left(1-\frac{2GM}{r}\right)dt^2-\frac{dr^2}{1-
\displaystyle\frac{2GM}{r}}-
r^2\left(d\theta^2+\sin^2\!\theta\,d\varphi^2\right),
\end{equation}
where the variable $r>0$ is defined as the internal curvature
radius of a closed homogeneous isotropic 2-space
$dY^2=\gamma_{ij}\, dy^i dy^j$, which does not depend on $y^i
\,(i=1,2)$, $G$ is the gravitational constant. The $Y$-space is
invariant under the group of motions $G_3$ (two translations that
allow moving to an arbitrary point $Y$, and a turn around a
point), and it can be reduced to the form $dY^2= r^2d\Omega$,
where $d\Omega\equiv d\theta^2 +\sin^2\!\theta d\varphi^2$ is the
unit 2-sphere $\mathbb{S}^2$,\, $y^i\!= (\theta,\varphi)$. As the
pole $\theta=0$ it is possible to take any of points
$\mathbb{S}^2$, and the turn around the pole is described by the
angle $\varphi\in [\,0,2\pi)$.

The space $X$ orthogonal to $Y$ is given in the Eulerian gauge, in
which one of the coordinates $x^I$ coincides with $r$.
Additionally, metric~(\ref{schwarz}) is independent of the
coordinate $t\in\mathbb{R}^1$. Thus the black hole in vacuum has
the group of motion $G_4$ acting on the hypersurface
$\mathbb{C}^3=\mathbb{R}^1\times\mathbb{S}^2$ (three translations
in $\mathbb{C}^3$ + rotation in $\mathbb{S}^2$).

Topologically geometry~(\ref{schwarz}) is a 4-cylinder with
homogeneous 3-surface $\mathbb{C}^3$ and radial coordinate $r$.
Its reduction to positive values, $r>0$, defines the Schwarzschild
sector of a black (or white) hole. Attempts to extend this
solution inevitably lead to regions occupied by matter. Therefore,
the problem of analytical continuation of metric~(\ref{schwarz})
must be solved using more general GR metrics with matter, which we
only restrict by requiring spherical symmetry $G_3$ in
$\mathbb{S}^2$.

Let us consider a class of such metrics, which in the orthogonal
gauge ($g_{Ii}=0$) have the form
\begin{equation}
\label{general-metrics} ds^2=dX^2-dY^2=-{\mathcal N}\! {\mathcal
K}dt^2+\frac{d\rho^2}{4\rho\mathcal{K}}-
\rho\left(d\theta^2+\sin^2\!\theta\,d\varphi^2\right),
\end{equation}
where the real variable $\rho^{-1}$ is not restricted by the sign
and is defined as the $Y$-space internal curvature scalar that
does not depend on $y^i$:
\begin{equation}
\label{R2} R_{ij}^{(Y)}=\rho^{-1}\gamma_{ij}\,,
\end{equation}
The 2-tensor Ricci is constructed from the metric $\gamma_{ij}$,
which by spherical symmetry can always be reduced to the form
$dY^2= \rho\, d\Omega$.

For general reference frames $\rho, \mathcal{K}, \mathcal{N}$ are
4-scalars depending on coordinates of the $X$-space
$dX^2=n_{IJ}dx^Ix^J$, where $\mathcal{K}$ is the kinetic term
$\rho$ and $\mathcal{N}$ is restricted by positive values:
\begin{equation}
\label{K}
\mathcal{K}=\frac{\rho_{,\mu}\rho^{,\mu}}{\!4\rho}\equiv-1-2\Phi\,,\qquad
\mathcal{N}\equiv N^2 > 0\,,
\end{equation}
$\Phi$ is the metric gravitational potential. The energy-momentum
tensor has the form $T_\mu^\nu={diag}(T_I^J,-p_\perp,-p_\perp)$,
where $T_I^J=(\epsilon+p) u_I u^J- p\delta_I^J$. At $\rho > 0$
functions $\epsilon, p, p_\perp$ and $u^\mu=(u^I,0,0)$ describe
the energy density, longitudinal and transversal tension, and
4-velocity of matter ($u_I u^I= 1$), respectively.
Metric~(\ref{general-metrics}) reduces to solution~(\ref{schwarz})
in the part of the domain $\sqrt\rho=r
>0$ free of matter. Hence, by analogy with the vacuum case and by definition, the domain $\mathcal{K}>0$
will be later referred to as the T-region of
space~(\ref{general-metrics}). There, in particular, the component
$T^{t}_{t}$ describes the pressure \footnote{In particular, under
the co-moving condition $T^{t}_{t}=-p$
in~(\ref{general-metrics}).}, and the meaning of other components
depends on the sign of $\rho$.

GR equations relate the metric and material scalars:
\begin{equation}
\label{phi} \Phi^\prime= 2\pi G P\,,\qquad \dot\Phi= 2\pi G
T_t^\rho\,,
\end{equation}
\begin{equation}
\label{N} \frac{N^\prime}{N}=\frac{2\pi
G\left(E+P\right)}{\mathcal K}\,,
\end{equation}
\begin{equation}
\label{perp} \frac{\left(\rho N E\right)^\prime}{N}+ p_\perp
-\frac{P}{2}=-\frac{\left(N^3T^{t\rho}\right)\dot{}}{4N^3\mathcal{K}}\,,
\qquad \dot{P}=\left(T_t^\rho\right)^\prime\,,
\end{equation}
where the prime and dot denote the partial derivative with respect
to $\rho$ and $t$, respectively,
\begin{equation}
\label{EP} P\equiv -T_t^t-\frac{\Phi}{4\pi G\rho}\,,\qquad E\equiv
T_\rho^\rho+\frac{\Phi}{4\pi G\rho}\,.
\end{equation}
By modeling the state of the effective matter, one can determine
$P$ and $E$ from the Bianchi identities~(\ref{perp}), and by
integrating~(\ref{phi}) and (\ref{N}) over $dx^I$ from the
external solution~(\ref{schwarz}) into the future one can
reconstruct the metric potentials. Similarly, in terms of the mass
function $m=m(x^I)$ equations~(\ref{phi}) read
\begin{equation}\label{mm}
\Phi\equiv-\frac{Gm}{\sqrt\rho}\,,\qquad m_{,I}=
4\pi\rho\,e_{IK}T^K_J\frac{\partial x^J}{N\partial t}\,,
\end{equation}
where $({}_{,I})\equiv\partial/\partial x^I$ and the absolute
antisymmetric tensor in $X$ has the form
\begin{equation}
\label{e} e_{IJ}=|\det\left(n_{IJ}\right)|^{1/2} \left(
\begin{smallmatrix}
0 & -1 \\
1 &  \;0
\end{smallmatrix}
\right),\qquad
e^{IJ}=|\det\left(n_{IJ}\right)|^{-1/2}\left(\begin{smallmatrix}
\;0 & 1 \\
-1 & 0
\end{smallmatrix}
\right).
\end{equation}

\section{Space-time near the integrable singularity}\label{space-time}

We are interested in solutions~(\ref{general-metrics}) generated
by metric~(\ref{schwarz}) that are geodesically complete. The
sufficient condition for the metric to pass through $\rho=0$ is
the finiteness of potentials $\Phi$ and $N$ and their derivatives.
We call such models black-white (or black/white) holes with
integrable singularity~\cite{ls}. They are described by the mean
(averaged over the vacuum state~(\ref{schwarz})) metrical
4-dimensional space without punctures, which provides a continuous
extension of affine parameters on world lines of test particles
through the singular hypersurface $\rho=0$. Using reference frames
constructed on these particles we investigate the geometry of
black-white holes outside the Schwarzschild sector.

The effective matter generated by strong gravitational field near
the singularity changes the space-time near $\rho=0$ and is the
physical reason for the existence of integrable singularity.   The
formation of matter in extremely strong gravitational fields is in
accordance with the Le Chatelier principle: this is how nature
reacts to a sharp increase in amplitudes of the metric potentials
at $\rho\rightarrow 0$ (which would continue to grow in the
absence of matter), thus, preventing them from divergence.
Specific quantum-gravitational mechanisms of mutual
transformations of material and gravitational degrees of freedom
under extreme conditions need to be considered separately. Here,
by developing papers~\cite{ls,*lsm}, we postulate the continuity
of the gravitational potentials~(\ref{general-metrics}) in
presence of the effective matter.

In the R-region of the space-time~(\ref{general-metrics}) with
signature $(+\,,-\,,-\,,-)$, specified by the generating
metric~(\ref{schwarz}), we have $\mathcal{K} < 0,\, \rho=r^2> 0$.
Here the hypersurface $r=0$ is degenerate and represents a
time-like world line of the center of a spherically symmetric
matter distribution of matter (for example, the center of a star).
If there is no matter in the R-region, $r > 2GM$ as follows
from~(\ref{schwarz}), i.e. the hypersurface $r=0$ does not lie in
the R-region.

In the T-region we have $\mathcal{K} > 0$ and the variable $\rho$
can have any sign because the physics in this region is determined
by non-linear quantum effects, and {\it a priori} there are no
grounds to believe that the signature remains
indefinite~\cite{sakharov}. Here the singular hypersurface
$\rho=0$ splits 4-space in domains with different signatures
$(-\,,+\,,-\,,-)$ for $\rho > 0$ and $(-\,,-\,,+\,,+)$ for $\rho <
0$. The complete geometry depends on the distribution and
properties of the effective matter near $\rho=0$.

We assume that the effective matter distribution maintains the
symmetry of the generating field. For example, the region that is
evolutionary adjacent to the Schwarzschild metric (for example,
$T_{\mu\nu}\neq 0$ for $r\le r_0=const< 2GM$ and $T_{\mu\nu}= 0$
at $r>r_0$) conserves Killing $t$-vector and depends only on
$\rho$. The region inside the star keeps the spherical symmetry
and the field homogeneity of the star. We will use these
constraints when constructing models in Sections
\ref{geodesics}-\ref{astrophysical}.

Several properties of geometries~(\ref{general-metrics}) should be
noted.
\begin{itemize}
\item{The continuity of the potential $\Phi$ implies integrability
of the function $P(t,\rho)$ along lines $t=const$
(see~(\ref{phi})):
\begin{equation}
\label{im} \Phi\!\left(t,\rho\right)=\Phi_0 +2\pi G \int_0 P
d\rho\,,\qquad \Phi_0=\Phi\!\left(t,0\right).
\end{equation}}
\item{The continuity of $\Phi$ and $N$ implies the integrability
$E(t,\rho)$ along lines $t=const$ (see~(\ref{N}))
\begin{equation}
\label{iN} N\!\left(t,\rho\right)= N_0\exp\left(2\pi G
\int_0\frac{E+P}{\mathcal K}d\rho\right),\qquad N_0=
N\!\left(t,0\right).
\end{equation}}
\item{For $\mathcal{K}_0\equiv-1-2\Phi_0= K_0^2=const > 0$ the
structure of space-time at $\rho\rightarrow 0$ has the form
\begin{equation}
\label{exp} ds^2=-d\tilde{t}^{\,2} +dUdV - UVK_0^2\sin^2\!\theta\,
d\varphi^2\,,
\end{equation}
\[
\tilde{t}= K_0\int N_0dt\,,\qquad \rho=K_0^2\,UV\,,\qquad
\theta=\frac{1}{2K_0}\ln\Big\vert\frac{U}{V}\Big\vert \,.
\]
The direction $\theta$ can be taken along any meridional direction
in $\mathbb{S}^2$ counted from arbitrarily chosen pole
$\theta=0$.}
\end{itemize}

Thus, the integrable singularity of a black-white hole includes
hypersurfaces-horizons $U=0$ and $V=0$ lying in the T-region and
intersecting along a space-like bifurcation line $U=V=0$.  They
separate the T-region in cone sectors of the black ($U<0, V<0$)
and white ($U>0, V>0$) holes and the static zone ($UV<0$). The
horizons have the cylindrical symmetry
$\mathbb{R}^1\times\mathbb{S}^2$ with the space-like longitudinal
axis $t\in\mathbb{R}^1$ and null geodesics in $\mathbb{S}^2$
($t=const$)\footnote{It is reminiscent of horizons $r=2GM$ of an
eternal black (white) hole in the Kruskal metric which are null in
the radial direction, space-like in $\mathbb{S}^2$ and intersect
on the space-like bifurcation 2-sphere.} Photons propagating in
these directions are kept gravitationally at $\rho=0$ by
infinitely oscillating in $Y$ within a finite interval of the
affine parameter. Trajectories of other particles, including
photons with projection in $\mathbb{R}^1$, intersect the singular
hypersurface $\rho=0$ and escape in other metric domains.

\section{Geometric maps of an oscillating hole}\label{geodesics}

The structure of a 4-space can be described by 2- and
3-dimensional cross-sections which can be covered with a net of
trajectories of test particles on them. It is convenient to choose
coordinates such that the light trajectories have a slope of
$\pi/4$ like in the flat world~\cite{penrose}. By spherical
symmetry all geodesics in~(\ref{general-metrics}) can be
represented by three groups of cross-sections:
\begin{itemize}
\item{$X$-planes - longitudinal cross-sections $(\theta,\varphi)=
const$;} \item{$UV$-planes - transversal cross-sections
$(t,\varphi)= const$;} \item{$tUV$-hypersurfaces  --
longitudinal-transversal cuts $\varphi= const$.}
\end{itemize}
The $X$-planes are filled with radial or longitudinal geodesics,
the $tUV$-planes contain particles propagating in both
longitudinal and transversal directions, and the $UV$-planes
orthogonal to the bifurcation line are filled with spiral
geodesics which do not escape beyond the $T$-region (see Appendix)
(Fig.~\ref{risunok1}).

\begin{figure}[t!]
\begin{center}
{\includegraphics{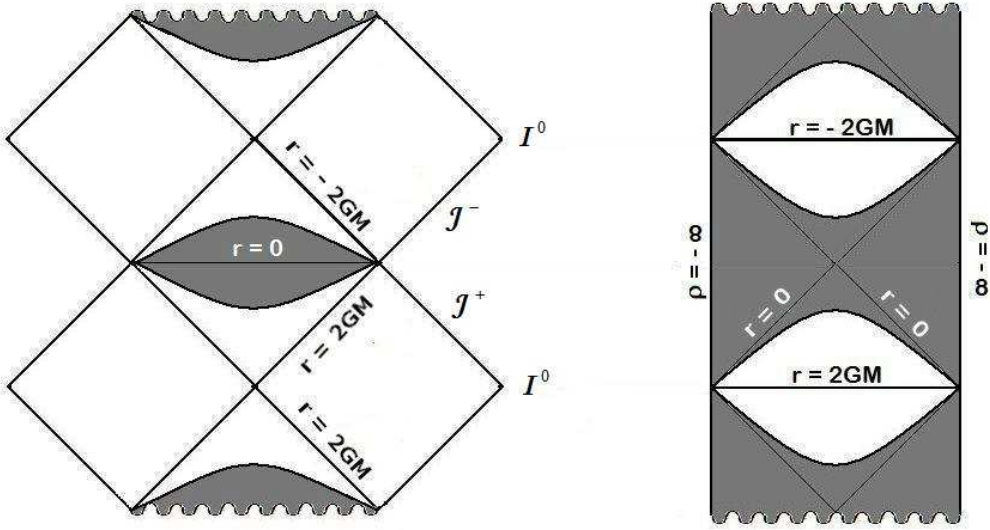}}
\end{center}
\caption{The longitudinal (a) and transversal (b) cross-sections
of an oscillating black-white hole. $\mathcal{J}^{+}$ is the light
infinity of the future for observers in the region $r>0$,
$\mathcal{J}^{-}$ is the light infinity of the past
in the region $r<0$, $I^{0}$ are the space infinities of R-regions.}%
\label{risunok1}
\end{figure}

As an example, consider metric~(\ref{general-metrics}),
independent of $t$, filled with a matter in the $T$-region $r\le
r_0=const<2GM$. The matter is distributed in the $UV$-planes along
bundles of straight lines $(t,\theta,\varphi)=const$ intersecting
at points $U=V=0$. Integrating~(\ref{mm}) along the time lines
$r=\sqrt{\rho}$ from the black hole region~(\ref{schwarz}) into
the future, we obtain continuous potentials
\begin{equation}
\label{m} \Phi=-\frac{Gm}{r}\,,\qquad
m=M-4\pi\int_{r_0}p(r)r^2dr=-4\pi\int_0 p\rho dr\,,
\end{equation}
which directly connect both holes. The black hole metric is in the
sector $r>0$, while the white hole is obtained from the extension
of solution~(\ref{m}) into region $r<0$ under the finiteness
condition $p\rho=\Phi_0/4\pi G=const$ at $\rho=0$. The complete
geometry can be restored by integrating from the bifurcation line
points along all bundles of lines, including regions $\rho < 0$.

For illustration, let us consider the model of an oscillating
black-white hole with triggered matter distribution depending on
$\rho\le\rho_0\equiv r_0^2$ (see~\cite{ls,lsm}, case B). In other
terms, the transversal pressure increases in a jump: $p_\perp=
\lambda_0\theta(\rho_0-\rho)$. The longitudinal pressure, which is
chosen to be vacuum-like ($p=-\epsilon$), is calculated using the
Bianchi identities~(\ref{perp}):
\begin{equation}
\label{mB} \rho\le\rho_0:\qquad P=-E=\frac 23 \lambda_0\,,\qquad
\Phi=\frac 12 H_1^2\left(\rho-3{\rho_0}\right),
\end{equation}
\begin{equation}
\label{sB} \rho >\rho_0:\qquad P=-E=\frac{M}{4\pi |r|^3}\,,\qquad
\Phi=-\frac{GM}{|r|}\,,\qquad
\end{equation}
where $H_1^2\equiv 8\pi G\lambda_0/3\equiv GM/r_0^3=const$.  By
substituting~(\ref{mB}) and~(\ref{sB})
into~(\ref{general-metrics}), we obtain the metric with matter
($\rho\le\rho_0$)
\begin{equation}
\label{obw} ds^2=-\mathcal{K}dt^2+\frac{d\rho^2}{4\rho\mathcal K}-
\rho d\Omega =\frac{-\left(1-
\tilde{\rho}\right)^2\!d\tilde{t}^{\,2}+
H_1^{-2}\tilde{\rho}^{-1}d\tilde{\rho}^2-
4r_1^2\tilde{\rho}d\Omega}{\left(1+\tilde{\rho}\right)^2}\,,
\end{equation}
and the metric in vacuum ($\rho=r^2>\rho_0$)
\begin{equation}
\label{schwa}
ds^2=\left(1-\frac{2GM}{|r|}\right)dt^2-\frac{dr^2}{1-
\displaystyle\frac{2GM}{|r|}}- r^2d\Omega\,,\qquad
\end{equation}
where $\tilde{t}=K_0t,\,
\rho=4r_1^2\tilde{\rho}(1+\tilde{\rho})^{-2},\, \mathcal{K}=K^2_0
- H_1^2\rho,\, r_1=K_0/H_1=r_0(3-r_0/GM)^{1/2},\,
K_0^2=-1-2\Phi_0,\, \Phi_0=\Phi(0)=-3GM/2r_0$.\, The value of the
potential of the integrable singularity is one and a half times
lower than that at the matter boundary, and this ratio does not
depend on the parameter $r_0$.

Let us represent~(\ref{obw}) using the proper interval on the axis
$\rho$:
\begin{equation}
\label{BW}\rho=r^2\in [0,\rho_0]:\quad
ds^2=d\tau^2-\cos^2\!\left(H_1\tau\right)d\tilde{t}^{\,2}
-r_1^2\sin^2\!\left(H_1\tau\right)d\Omega,\;
\end{equation}
\begin{equation}
\label{Bif}\rho\le 0:\qquad
ds^2=-dx^2-\cosh^2\!\left(H_1x\right)d\tilde{t}^{\,2}
+r_1^2\sinh^2\!\left(H_1x\right)d\Omega\,,
\end{equation}
where  $r=-r_1\sin(H_1\tau)\in[-r_0,r_0]$ and
$\rho=-r_1^2\sinh^2(H_1x)\le 0$. The space-time of static domains
is asymptotically the anti-de-Sitter one:
\begin{equation}
\label{ADS}\rho\ll-r^2_0:\qquad\qquad
T_\nu^\mu=-\lambda_0\delta_\nu^\mu\,,\qquad\quad
R_\nu^\mu=3H_1^2\delta_\nu^\mu\,.\qquad
\end{equation}
For this reason the regions $\rho<0$ will be also referred to as
anti-de-Sitter zones (ADS).

\section{The source of eternal black (white) hole}\label{black-white}

Eternal holes, in which matter is absent almost everywhere, can be
obtained in the limit of small $r_0\rightarrow 0$ in
equations~(\ref{mB})-(\ref{schwa}):
\begin{equation}
\label{schwat} r\in\mathbb{R}^1:\qquad\qquad
\epsilon=-p=2p_\perp=M\,\frac{\delta\!\left(r\right)}{2\pi
r^2}\,,\quad
\end{equation}
\begin{equation}
\label{schwatt} \rho < 0:\qquad\qquad p=-\epsilon=p_\perp
=\lambda_0\equiv\frac{3M}{8\pi r_0^3}\,.\quad
\end{equation}
The ADS zones are strongly curved by the dense vacuum $\lambda_0$
(see (\ref{Bif})):
\begin{equation}
\label{ads2} \rho\ll-r_0^2:\qquad ds^2=-dx^2+\frac 34
r^2_0e^{2H_1x}\left(-H^2_1dt^2 +d\theta^2+\sin^2\!\theta
d\varphi^2\right),
\end{equation}
which produces a $\delta$-like material source of geometry of the
eternal black hole~(\ref{schwa}) in the region $r\in\mathbb{R}^1$.
This source localized at $|r|\le r_0$ has the density
$\sim\lambda_0$ and total mass $M$. The value $r_0$ can be
calculated in quantum field theory (see an estimate
in~\cite{ls,*lsm}). In our classical treatment $r_0$ is a free
parameter of the problem. Note also that the relation between the
longitudinal and transversal tensions in~(\ref{schwat}) is not
universal, it is related to the chosen condition of the model
$p=-\epsilon$.

Solution (\ref{schwat})-(\ref{ads2}) shows that the polarized
vacuum in static ADS zones is the source of eternal black holes.
The gravitational mass of the effective matter of each of the
ADS-zones generates, via its light hypersurface of the future
$\rho=0$, the causally connected Schwarzschild metric of a
black-white hole in vacuum, which extends in time up to the next
integrable singularity $\rho=0$, where the process repeats (see
Fig.~\ref{risunok1}). The full geometry is invariant with respect
to the inversion $r\rightarrow -r$ of the spherical coordinate
system relative to the bifurcation lines. In this sense phase
transitions between gravitational and material degrees of freedom
in this model are reversible.

In Section \ref{astrophysical} we construct one more example of
reversible geometry, in which the black-hole source is a parent
star that collapses from the R-region with the formation of a
singularity in the T-region. In the course of evolution the
space--time passes the stage of effective matter and transforms
into a white hole with the metric of a homogeneous cosmological
model.

\section{Astrophysical black-white hole}\label{astrophysical}

Let us consider the model of a black-white hole with integrable
singularity assuming that the black hole was formed during the
collapse of a star from the parent universe. The star is modeled
as a homogeneous sphere with radius $r(T,R=1)=a(T)$, which was at
rest in a flat space-time and had the mass $M$ ($a\gg 2GM$ at
$T\rightarrow -\infty$) and then started collapsing due to
self-gravitation, with initial pressure being negligible. Here $T$
and $R$ are the Lagrange coordinates co-moving with the star
material, and the proper time $T\in\mathbb{R}^1$ and radial
markers of spherical shells $R\ge 0$ are normalized such that
$R=1$ at the star's surface.

The field symmetry is determined by the initial conditions of the
problem. Inside the star ($R\le 1$) the symmetry
$\mathbb{R}^1\times\mathbb{E}^3$ with 6-parametric group $G_6$ on
$\mathbb{E}^3$ and potentials $a, H, \epsilon, p$ are functions of
$T$. Outside the star ($R > 1$) the group of motion $G^4$ on
$\mathbb{C}^3$ and all potentials depend on $r$. To avoid crossing
of matter flows we assume that the longitudinal tension, which is
produced outside the star at large curvatures, is vacuum-like.
Then at $R > 1$, $N=1$ and the energy-momentum tensor is invariant
with respect to motions in $X$: $T_\nu^\mu =
-{diag}(p,p,p_\perp,p_\perp)$. Inside $R\le 1$ the matter is
Pascalian, $T_\nu^\mu=(\epsilon+p) u_\nu u^\mu - p\delta_\nu^\mu$
(where 4-velocity $u_\nu=T_{,\nu}$), and its state is calculated
from boundary conditions. The homogeneity of the star suggests the
continuity of $\Phi$ and $p$ at the boundary $R = 1$ ($r=a$),
whereas the density and transversal pressure can change
discontinuously.

Inside the star the metric has the form~\cite{lm}
\begin{equation}
\label{shar} ds^2=dT^2-a^2\!\left(dR^2+R^2d\Omega\right)=
\frac{a^4\!H^2dt^2- dr^2}{1+2\Phi}-r^2d\Omega\,,
\end{equation}
where the Euler and Lagrange coordinates are related as
\begin{equation}
\label{rR} R\le 1:\qquad\quad r=aR\,,\qquad\;
t=-\int\frac{dT}{a^2H}-\frac{R^2}{2}\,,\quad
\end{equation}
and the Hubble function $H=da/adT$ can be found from the Friedmann
equations:
\begin{equation}
\label{H} H^2=\frac{8\pi
G}{3}\epsilon=\frac{2Gm}{r^3}=-\frac{2\Phi}{r^2}\,,\qquad
\frac{d\epsilon}{dT}+3H\left(\epsilon+p\right)=0\,.
\end{equation}
Initially the pressure is absent and the star collapses freely
($H<0,\, a^3\!H^2=2GM$). The Newtonian potential in this limit is
$\Phi_N=GM(R^2-3)/2a$. The internal tension in the star appears at
$a\le r_0 < 2GM$, and it can be calculated by sewing it to the
effective matter at the boundary $R=1$.

Outside the star the metric has the form
\begin{equation}
\label{sharv} ds^2=\left(1+2\Phi\right)\!dt^2-
\frac{dr^2}{1+2\Phi}-r^2d\Omega= dT^2+ 2\Phi dR^2- r^2d\Omega\,,
\end{equation}
where the Lagrange reference frame co-moves with shells of free
dust particles following the collapse of the star surface:
\begin{equation}
\label{tT}  R> 1:\qquad R - T
=\int\frac{dr}{\sqrt{-2\Phi}}\,,\qquad\; t = T
-\int\frac{\sqrt{-2\Phi}dr}{1+2\Phi}\,,
\end{equation}
\begin{equation}
\label{tT} \Phi=-\frac{Gm}{r}\equiv-\frac 12 H^2r^2,\,\qquad
m=M-4\pi\int_{r_0}p\!\left(r\right)r^2dr\,.
\end{equation}
The source of the Schwarzschild metric is the mass $M$ of the star
with the continuous potential $\Phi=-GM/r$ at the star boundary.
The discontinuity of functions $N,\, t$ and $g_{RR}$ is due to the
density jump at $R=1$. The effective matter outside the star
emerges at $r\le r_0$ and preserves the symmetry of the parent
metric. The continuity of $\Phi$ suggests $m(0)=0$ and leads to
formulas of Section \ref{geodesics}.

\begin{figure}[t!]
\begin{center}
{\includegraphics{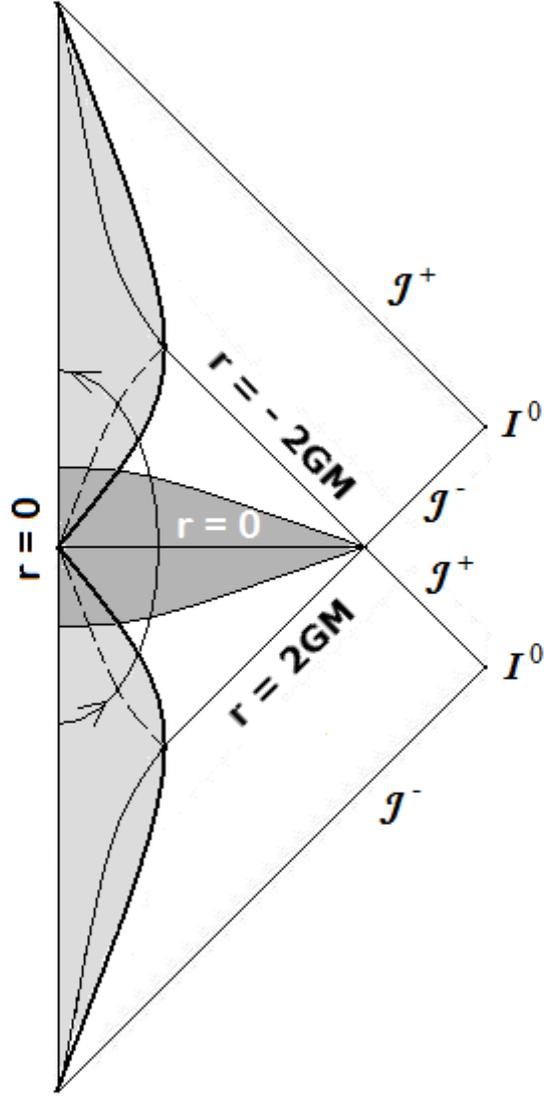}}
\end{center}
\caption{The Penrose diagram of an astrophysical black-white hole.
The hell gray region is the body of the star, the dark gray region
is the effective matter, the dashed line separates the R- and
T-regions inside the star, the line with arrows is the contour $t=const$.}%
\label{risunok2}
\end{figure}

By extending the reference frame of free particles over the entire
cone $\rho=r^2\ge 0$, we continue metric from the black hole
region ($r > 0$) into the white hole region ($r < 0$) bypassing
the body of the star itself ($R>1$) (see (\ref{m})-(\ref{schwa})).
Closing the external metric, we then restore the complete solution
on the entire manifold $R\ge0$, by sewing $p$ at the star boundary
(Fig.~\ref{risunok2}).  As a result, everywhere in the effective
matter zone we obtain $P=2\lambda_0/3$ and
\begin{equation}
\label{vnu} |r|\le r_0 \tilde{R}:\qquad
r=-\sqrt{3}r_0\tilde{R}\sin\!\left(H_1\tilde{T}\right),\quad
\Phi=\frac 32\Phi_0\tilde{R}^2\cos^2\!\left(H_1\tilde{T}\right),\;
\end{equation}
where $\tilde{R}\equiv\min(R,1),\; \tilde{T}\equiv
T+(1-R)\cdot\theta(R-1)$.\, The state of effective matter inside
the star is described by the equation $\epsilon +
3p=2\lambda_0$\,.

Freely falling shells of the star intersect simultaneously in one
point $r=0$ separating the time-like and space-like parts of the
axis $r=0$ (see Fig.~\ref{risunok2}). At this bifurcation point
the equation of state of the effective matter is $p=-\epsilon/3$
and the motion is rectilinear: $a\propto T$. Gravitation is
'switched off' due to the zero mass at $r=0$, which provides
finite amplitude of the tidal forces and the continuation of world
lines of the matter into future.

\section{Physical nature of integrable singularities}\label{nature}

The concept of integrable singularities allowed us to advance in
solving the cosmogenesis problem using the new class of
black-white hole models in GR.  Are these solutions only research
tools or they really exist? If yes, then why and how do they form?
These questions require separate studies. Here we discuss some
ideas on the physical nature of integrable singularities.

Reasons for the formation of singular structures with finite
gravitational potential are related to the redistribution of
material and space-time degrees of freedom in strong gravitational
fields of singularities themselves. (At high intensities and small
scales quantum effects apparently play the key role). That is why
we speak of the effective matter, which includes both material and
gravitational degrees of freedom that remain after the metric
averaging over the quantum state. However, we keep notions of
energy-momentum and laws of motion in the form of the Bianchi
identities. This allows us to describe the back reaction using the
Einstein equations in which the left side is defined using the
mean metric $g_{\mu\nu}$, and the right side contains the
energy-momentum tensor of the effective matter $T_\nu^\mu$
including all polarization, gravitationally modified and other
terms of quantum theory that need to be calculated.

Let us illustrate the method by considering the behavior of one
degree of freedom in the given physical symmetry of the collapsing
star. Let the physical variable be described by a massless field
$\varphi$ minimally coupled to metric~(\ref{shar}).  The second
quanization leads to the following equation for amplitudes of
Fourier harmonics $\varphi_k=\nu_k(\eta)/a$ \cite{lm}:
\begin{equation}
\label{ud} \nu^{\prime\prime}_k+\left(k^2- \bar{U}\right)\nu=0\,,
\end{equation}
\begin{equation}
\label{U}
\bar{U}=\bar{U}(\eta)=\frac{a^{\prime\prime}}{a}=a^2H^2+\!\left(aH\right)^\prime
=-2\Phi+\frac{\Phi^\prime}{\sqrt{-2\Phi}}\,,
\end{equation}
here the prime denotes the derivative with respect to the
conformal time $\eta=\int dT/a$, $k$ is the wave number, $\Phi$ is
the gravitational potential on the stellar surface.

As long as the pressure in the star is low, the tidal potential
increases with decreasing radius: $\bar{U}=GM/a=a^2H^2/2$. At
large $a$ the state of the field is vacuum-like: it oscillates
near an equilibrium point in the adiabatic zone $\bar{U} < k^2$.
During the collapse the field enters the parametric zone, and its
amplitude increases unlimitedly as $U > k^2$ increases:
\begin{equation}
\label{n}
\varphi_k=\frac{\exp\!\left(-ik\eta\right)}{a\sqrt{2k}}\rightarrow\frac{H}{k^{3/2}}\,,\qquad\;
\varphi_{,\mu}\varphi^{,\mu}\simeq H^4\ln \bar{U}\,.
\end{equation}
At this stage, the pressure cannot be neglected any more, vacuum
is polarized and, in its turn, affects the metric, which changes
the gravitational potential and the rate of collapse near $a=0$
(for the solar mass and the number of degrees of freedom $\sim
100$ this occurs at $H\simeq 0,1\times M_P(\ln \bar{U}_0)^{-1/2}
\sim 10^{17}$~GeV).

One can suppose that the back reaction, described by the Friedmann
equations, restructures the solution in such a way that tidal
potential (\ref{U}) stops increasing and saturates:
$\bar{U}(r\rightarrow 0)=\bar{U}_0=-2\Phi_0=const$. This precludes
the ultraviolet catastrophe and saves oscillators from being
destroyed, because high-frequency modes with $k^2 > U_0$ remain in
the adiabatic zone and are not polarized. Thus the notion of the
Heisenberg state vector that stems from the Minkowski vacuum in
the R-region stays valid, and the main requirement of the metric
theory on the finiteness of the gravitational potential on the
singular hypersurface is secured. We call such a structure an
integrable singularity.

This example illustrates the difference of our theory from models
with bounces (see, for example,~\cite{Nov66}), which have no
singularity, $a > 0$, and at the bounce $H=\Phi=0$.  The last
requirement, in our opinion, is redundant. In our models the
potential $\Phi$ achieves extremum at the singularity $r=0$, which
weakens the singularity but does not fully eliminate it. For small
$a\propto T$ the situation resembles the explosive model of
non-gravitating particles by Milne. At the moment of crossing,
particles move by inertia and do not feel the central mass
attraction. Gravitation at this moment is switched off ($m(0)=0$),
therefore the space does not bend, although the energy density
diverges.

\section{The arrow of time and natural selection}\label{arrow}

Gravitationally tidal models of black-white holes containing
expanding flows of matter elucidate many fundamental unanswered
questions in the modern physics. One of them is the causality
principle, which puts the causal relationships in correspondence
with the arrow of time. In the context of our work we can discuss
the origin of the cosmological arrow of time, which we consider as
the orientation of the future light cone in the direction of the
volume expansion of a large-scale flow of matter.

As is well known, dynamical equations describing microscopic
processes are invariant relative to the change of the sign of
time. However, the local dynamics has to be completed with an
external arrow of time, since the invariance relative to the
direction of time is lost in the limit transition to the global
geometry.

Geodesically full geometries with integrable singularities suggest
the origin of the arrow of time. They include different space-time
domains separated by event horizons $r=0$ and $r=2GM$:
non-stationary regions of black and white holes (lying between the
Schwarzschild and singular horizons)  and  alternating static
R-zones (connecting the Schwarzschild horizons of white and black
holes), and ADS-zones (connecting singular horizons of black and
white holes).  Here all possibilities are realized.  Each
collapsing and anti-collapsing (cosmological) region has its own
time. The static regions are time-independent but are not
spatially homogeneous.

We deal with a unique geometry that is split into sectors with
time and space parity. When crossing any of the horizons the sense
of the coordinate $r$, which the metric depends on,
changes~\cite{Nov61}. In some domains $r$ is the time coordinate
(and then we obtain black holes and/or cosmological models), in
other domains it is the space coordinate (static zones). The
complete geometry here can remain invariant relative to the change
$r\rightarrow -r$.

Therefore, the origin of the cosmological arrow of time is related
to the initial conditions. We (observers) belong to a cosmological
flow of matter and live in its proper time which started at the
moment $r=0$ some 14 billion years ago. The time coordinate can be
extrapolated into the past, to the pre-cosmological epoch, but
there it described the time on the T-zone of the parent black
hole. Furthermore, this coordinate represented the radial
coordinate of an asymptotically flat space of the parent universe,
in which the star had lived before it collapsed to the black hole.
The integrable singularity $r=0$ that emerged during the collapse
'kindled' our Universe, and after several billion years the
non-linear large-scale structure began developing to initiate the
process of star formation. As a result, new black holes appeared;
they can be entrances to new universes.

This process of evolution of the multisheet space-time resembles
the growth of a tree (a genealogical tree, so to speak). Such a
tree can flourish, or can wither if no new black holes are formed
in the daughter universes.  The critical situation appears when
the development conditions do not provide the production of seed
density fluctuations to form gravitationally bound matter clumps
and their collapse to black holes. But another scenario can be
realized: one collapse under favorable conditions, which realize
inflationary parameters and phase transitions in a white hole, can
lead to flourishing of a whole tree with non-decaying chains of
new universes. Because of these processes the cosmological natural
selection works~\cite{LSmolin99}: only universes where black holes
can be formed survive and develop, and this is possible for a
certain set of parameters, world constants, etc.

This concept of multisheet hyperverse is based on the
gravitational instability processes which resemble oscillating
tides. Anti-collapsing space-time regions (white holes) stem from
collapsing black holes, and, inversely, an expanding
quasi-homogeneous flow of matter of a white hole desintegrates
into clumps collapsing into black holes. The former process is
related to the $r=0$ horizon, while the latter - to the $r=2GM$
horizon. At both horizons the gravitational potential is
relativistic, and quantum-gravitational processes of vacuum
polarization and pair creation should be taken into account.
However, while on the Schwarzschild horizon these effects are
suppressed by the mass parameter (the Hawking evaporation), at
$r=0$ they dominate and form structures of integrable
singularities.

\section{Conclusions}

In the framework of our concept of integrable singularities a new
class of black-white holes in GR is obtained, which can be the key
to solve the cosmogenesis problem. The integrable singularity
$r=0$ can be compared with a classical cusp in which the energy
density or the longitudinal tension of matter diverge, but the
mass is zero \footnote{The time-like axis of the zero mass (the
line of the star center) lies in the $R$-region, and the mass
function can be found by integrating the density over radial
shells (see~(\ref{mm})): $m=4\pi\int_0\epsilon(r)r^2dr$. The
space-like axis of the zero mass (bifurcation lines
(see~(\ref{exp})) lies in the $T$-region, and the mass function of
a black-white hole can be obtained by integrating the longitudinal
tension over time: $m=-4\pi\int_0p(r)r^2dr$.}, $m(0)=0$, and the
gravitational potential $\Phi_0$ is limited. Because of this
property the tidal forces are finite and the geodesics freely
extend from a black hole into a white hole, where the geometry is
equivalent to that of an expanding cosmological model.

The mass of matter in the multisheet hyperverse can be arbitrarily
large, since it is compensated by the negative gravitational bound
energy. Thus the total energy of holes, measured in static zones,
is constant in time. The integrable singularities is reminiscent
of machines for reprocessing gravitational degrees of freedom into
material ones, however quantitative characteristics of this
process can be determined only in the self-consistent quantum
theory.

\begin{acknowledgments}
This work is supported by Ministry of Education and Science of the
Russian Federation (contract no. 16.740.11.0460 of 13.05.2011), by
the Russian Foundation for Basic Research (OFI 11-02-12168 and
12-02-00276) and by grant no. NSh 2915.2012.2 from the President
of Russia. VNS thanks FAPEMIG for support.
\end{acknowledgments}

\appendix*

\section{Motion of test particles in a black-white hole}

\begin{figure}[t!]
\begin{center}
{\includegraphics{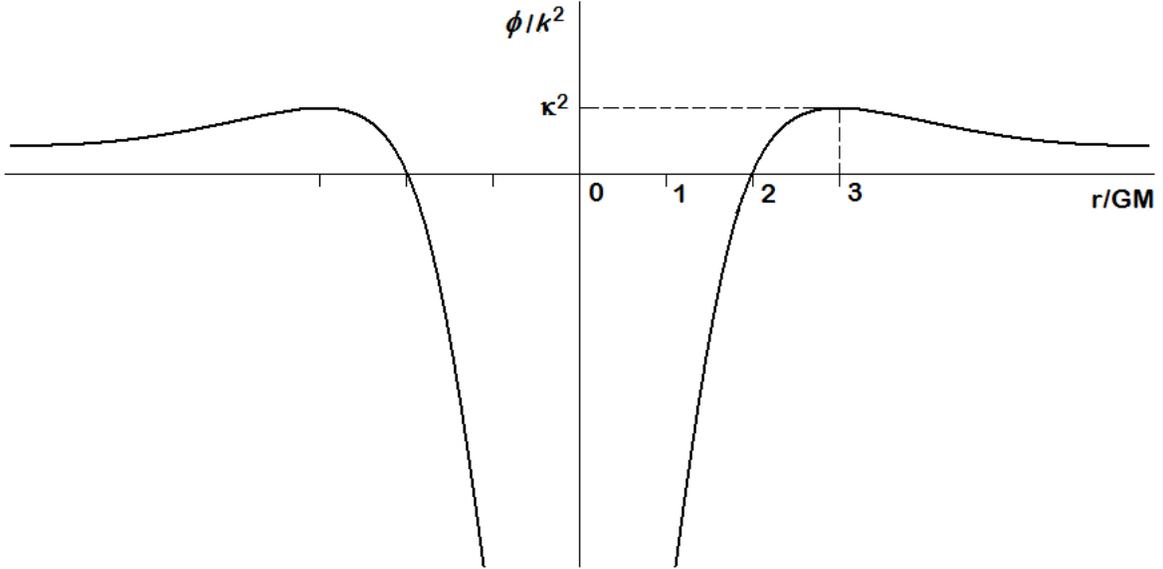}}
\end{center}
\caption{The potential $\phi(r)$ of light geodesics in a black-white hole.}%
\label{risunok3}
\end{figure}

Consider the motion of test particles in
metric~(\ref{general-metrics}), which is independent of coordinate
$t$. Let $k^\mu= dx^\mu/d\lambda$ and $\lambda$ be the tangent
vector, and the affine parameter along the trajectory $x^\mu=x^\mu
(\lambda)$. From the geodesic equation $k^\mu k_{\nu;\mu} =0$ we
obtain
\begin{equation}
\label{id} k\equiv k_t=const\,,\qquad k_\perp\equiv k_\theta +
\frac{k_\varphi}{\sin^2\!\theta}=const\,,
\end{equation}
where $k_\varphi=const$ is the azimuthal angular momentum. Unlike
invariants of motion for the longitudinal and transversal momenta,
the value of the azimuthal number depends on the orientation of
the polar coordinate system. By adjusting the pole $\theta=0$ with
one of the points on the trajectory, we have $\varphi=const$ and
$k_\varphi=0$ everywhere on the world line of a particle.  In the
projection onto a 2-sphere the particle moves along a meridian
with monotonic increase of the angle $\theta$ within the interval
$2\pi$ by the number of turns in $\mathbb{S}^2$:
\begin{equation}
\label{thet}
\frac{d\theta}{d\lambda}=-\frac{k_\perp}{\rho}\,,\qquad
\left(\frac{d\rho}{2d\lambda}\right)^2-
\left(k_\perp^2+n\rho\right)\mathcal{K}=\frac{k^2\rho}{N^2}\,,
\end{equation}
The equation for $\lambda(\rho)$ follows from the normalization
integral $k_\mu k^\mu\equiv n =const$, where $n=0$ for the null
and $n=1$ for the time-like geodesics, respectively. Inside the
cone $\rho=r^2\ge 0$ of a black-white hole equation~(\ref{thet})
has the form:
\begin{equation}
\label{lam+} \left(\frac{dr}{d\lambda}\right)^2+\phi
=\frac{k^2}{N^2}-n\,,\qquad
\phi=\frac{k_\perp^2}{r^2}\left(1+2\Phi\right)+2n\Phi\,,
\end{equation}
where the potential $\phi=\phi(r)$ tends to zero at
$r\rightarrow\infty$ (Fig.~\ref{risunok3}).

The longitudinal light ($n=k_\perp=0$) geodesics propagate
directly from the black hole into the white hole with continuous
affine time $\lambda=-k^{-1}\int Ndr$. The spiral light geodesics
on the cross-cuts ($n=k=0$) lie in the T-region of the black or
white hole, $\mathcal K=K^2\ge 0$, coming from ADS-zones and going
back to ADS-zones: \, $\theta=-k_\perp\int d\lambda/\rho=\pm \int
d\rho/(2\rho K)$. Of special interest are photons with
$n=k=k_\perp=0$ living on the horizons $\rho=0$:
\begin{equation}
\label{ps}
\theta=\pm\,\frac{\ln|\lambda-\lambda_0|}{2K_0}\,,\qquad
\rho=\mp\,2K_0\left(\lambda-\lambda_0\right)k_\perp \rightarrow 0,
\end{equation}
where $\lambda_0$ is the value of the affine parameter on the
bifurcation line,

Trajectories of photons with the impact parameter $\kappa\equiv
k_\perp (3\sqrt{3}GMk)^{-1}\!< 1$ connect both R- and T-regions.
Photons with $\kappa=1$ are turning in R-regions at radius
$r=3GM$. For $\kappa>1$ there are photons of two kinds: in the
R-region with $r>3GM$, and in the zone $r < 3GM$ which unites the
T-region and the inner part of the R-region adjacent to it.

\newpage
\bibliography{arxivUFN2}

\end{document}